\documentclass[]{spie}

\usepackage[]{graphicx,color}
\usepackage[]{amsmath,amssymb}

\newcommand{\farcs}{\hbox{$.\!\!^{\prime\prime}$}}

\title{MagAO: Status and on-sky performance of the Magellan adaptive optics system}

\author{
Katie M. \mbox{Morzinski\authorinfo{\supit{*}NASA Sagan Fellow.  Contact KMM at ktmorz@arizona.edu}\supit{*a},}
Laird M. \mbox{Close\supit{a},}
Jared R. \mbox{Males\supit{*a},}
Derek \mbox{Kopon\supit{b},}
Phil M. \mbox{Hinz\supit{a},}
Simone \mbox{Esposito\supit{c},}
Armando \mbox{Riccardi\supit{c},}
Alfio \mbox{Puglisi\supit{c},}
Enrico \mbox{Pinna\supit{c},}
Runa \mbox{Briguglio\supit{c},}
Marco \mbox{Xompero\supit{c},}
Fernando Quir{\'o}s-\mbox{Pacheco\supit{c},}
Vanessa \mbox{Bailey\supit{a},}
Katherine B. \mbox{Follette\supit{a},}
T. J. \mbox{Rodigas\supit{d},}
Ya-Lin \mbox{Wu\supit{a},}
Carmelo \mbox{Arcidiacono\supit{e},}
Javier \mbox{Argomedo\supit{f},}
Lorenzo \mbox{Busoni\supit{c},}
Tyson \mbox{Hare\supit{g},}
Alan \mbox{Uomoto\supit{g},}
and
Alycia \mbox{Weinberger\supit{d}}
\skiplinehalf
\mbox{\supit{a} CAAO}, Steward Observatory, University of Arizona, Tucson AZ 85721, USA;
\mbox{\supit{b} Max}-Planck-Institut f{\"u}r Astronomie, K{\"o}nigstuhl 17, D-69117 Heidelberg, Germany;
\mbox{\supit{c} INAF}-Osservatorio Astrofisico di Arcetri, Largo E. Fermi 5, I-50125 Firenze, Italy;
\mbox{\supit{d} Department} of Terrestrial Magnetism, Carnegie Institute of Washington, 5241 Broad Branch Road NW, Washington, DC 20015, USA;
\mbox{\supit{e} INAF}-Osservatorio Astronomico di Bologna, Via Ranzani 1, I-40127  Bologna, Italy;
\mbox{\supit{f} European} Southern Observatory, Karl-Schwarzschild-Str.\ 2, D-85748 Garching bei M{\"u}nchen, Germany;
\mbox{\supit{g} Observatories} of the Carnegie Institution of Washington, 813 Santa Barbara St. Pasadena, CA 91101, USA
}

\begin{document} 
\maketitle

\begin{abstract}
MagAO is the new adaptive optics system with visible-light and infrared science cameras, located on the 6.5-m Magellan ``Clay'' telescope at Las Campanas Observatory, Chile.
The instrument locks on natural guide stars (NGS) from 0$^\mathrm{th}$ to 16$^\mathrm{th}$ $R$-band magnitude, measures turbulence with a modulating pyramid wavefront sensor binnable from 28x28 to 7x7 subapertures, and uses a 585-actuator adaptive secondary mirror (ASM) to provide flat wavefronts to the two science cameras.
MagAO is a mutated clone of the similar AO systems at the Large Binocular Telescope (LBT) at Mt.\ Graham, Arizona.
The high-level AO loop controls up to 378 modes and operates at frame rates up to 1000 Hz.
The instrument has two science cameras: VisAO operating from 0.5--1 $\mu$m and Clio2 operating from 1--5 $\mu$m.
MagAO was installed in 2012 and successfully completed two commissioning runs in 2012--2013.
In April 2014 we had our first science run that was open to the general Magellan community.
Observers from Arizona, Carnegie, Australia, Harvard, MIT, Michigan, and Chile took observations in collaboration with the MagAO instrument team.
Here we describe the MagAO instrument, describe our on-sky performance, and report our status as of summer 2014.
\end{abstract}

\section{Introduction}
The twin Magellan telescopes feature two 6.5-m paraboloid primary mirrors that were spin-cast at the Steward Observatory Mirror Lab at the University of Arizona.  They were both dedicated in December 2000 at Las Campanas Observatory (LCO), Chile, and science operations began in 2001\cite{shectman2003}.  The telescopes are operated by a consortium from the Observatories of the Carnegie Institution of Washington (OCIW), the University of Arizona (UA), Harvard University, the Massachusetts Institute of Technology (MIT), and the University of Michigan, and are designed for best optical quality at an excellent site with median seeing of 0\farcs65 in $V$-band\cite{thomasosip2008,floyd2010}.  After over a decade of successful operations of the Magellan telescopes\cite{osip2008}, the first adaptive optics (AO) system has been installed on the Magellan Clay telescope at the bent Gregorian focus on the Nasmyth platform.  This instrument is called ``MagAO'' (we say ``Mag'' with a hard G as in ``Tag'').  MagAO is a UA instrument led by P.I.\ Laird Close, with contributions from our research partners INAF--Osservatorio Astrofisico di Arcetri of Italy and OCIW--Pasadena, and industrial partners Microgate S.r.l.\ and ADS International S.r.l.\ of Italy.

\section{Magellan adaptive optics}
\begin{figure}[!h]
	\centering
	\includegraphics[width=0.8\linewidth]{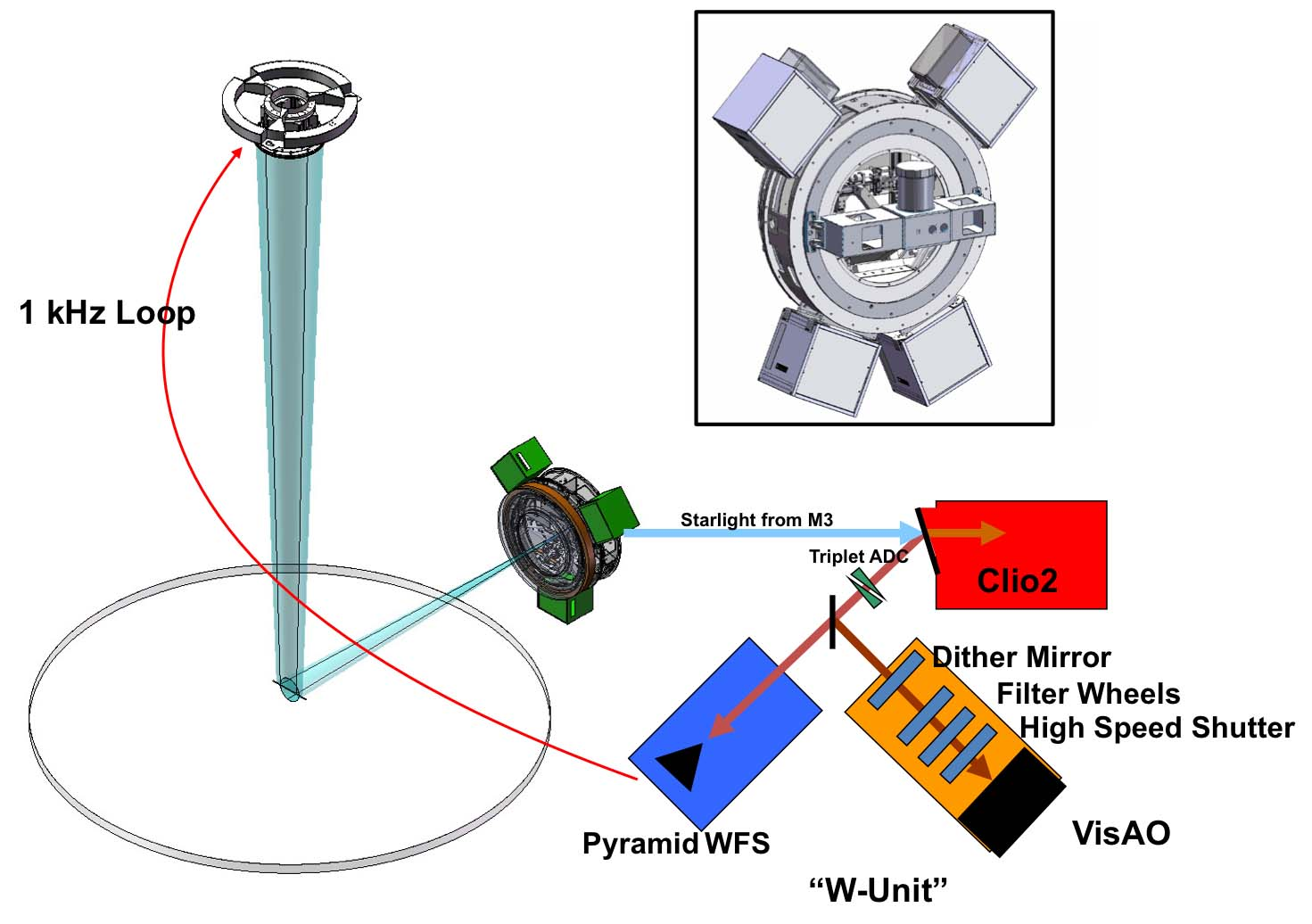}
	\caption{\label{fig:magao}
	Sketch of the MagAO instrument.
	The circle at lower left denotes the 6.5-m-diameter primary mirror of the Clay telescope pointed at zenith.
	The 85.1-cm-diameter adaptive secondary mirror (ASM) hangs above the primary, and sends light into the pyramid wavefront sensor (PWFS) and science cameras at the folded Nasmyth port (NAS) at the bent Gregorian focus.
	Within the NAS ring is the W-unit (including the PWFS and VisAO), while Clio2 is mounted externally.
	The exploded view at lower right shows that the starlight from M3, after it enters the NAS unit, is split between the IR science camera Clio2, the visible science camera VisAO, and the Pyramid WFS.
	Inset at upper right: CAD drawing of the NAS with the W-unit and Clio2.
	The instrument as built contains 4 electronics boxes mounted on the outer edge of the NAS ring as shown in the CAD drawing, not the 3 shown in green.
	The entire NAS ring rotates as the telescope tracks, with the full 360 degrees allowed.
	}
\end{figure}

MagAO was designed as a modification of the LBT AO systems FLAO and LBTI\cite{esposito2011spie}, and tailored for visible AO.\cite{close2008spie,close2010spie,close2012spie}
A sketch of the entire MagAO system is shown in Fig.~\ref{fig:magao}.
Here we describe the optical, mechanical, and software portions of the AO system.

\subsection{Pyramid wavefront sensor}
The pyramid wavefront sensor (PWFS) was first proposed for astronomical adaptive optics by Ragazzoni (1996)\cite{ragazzoni1996}, who noted the ability of the PWFS to operate in a wide range of seeing conditions and natural guide star (NGS) magnitudes due to its inherent flexibility.
The pyramid benefits from closed-loop operation in which wavefront compensation provides a reduced reference spot size;
in this condition PWFS has an enhanced sensitivity with respect to an equivalently spatial-sampled Shack-Hartmann sensor\cite{ragazzoni1999}.
We realize the pyramid flexibility with MagAO through binning the pixels on the WFS CCD, modulating the spot on the tip of the pyramid at different amplitudes, and sampling the wavefront at faster (slower) frame rates and gains for brighter (fainter) NGS magnitudes (respectively).

\begin{figure}[!h]
    \centering
    \includegraphics[width=0.4\linewidth]{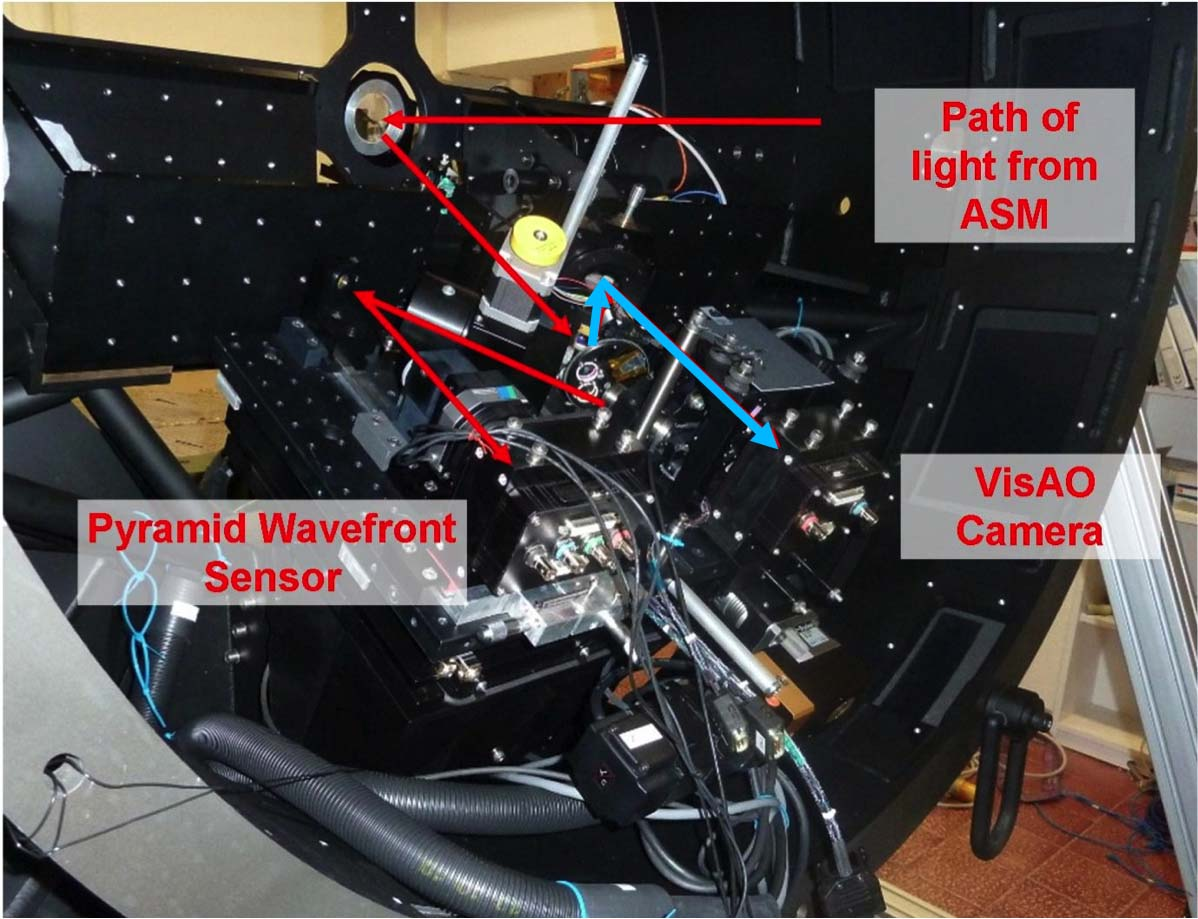}
    \includegraphics[width=0.59\linewidth]{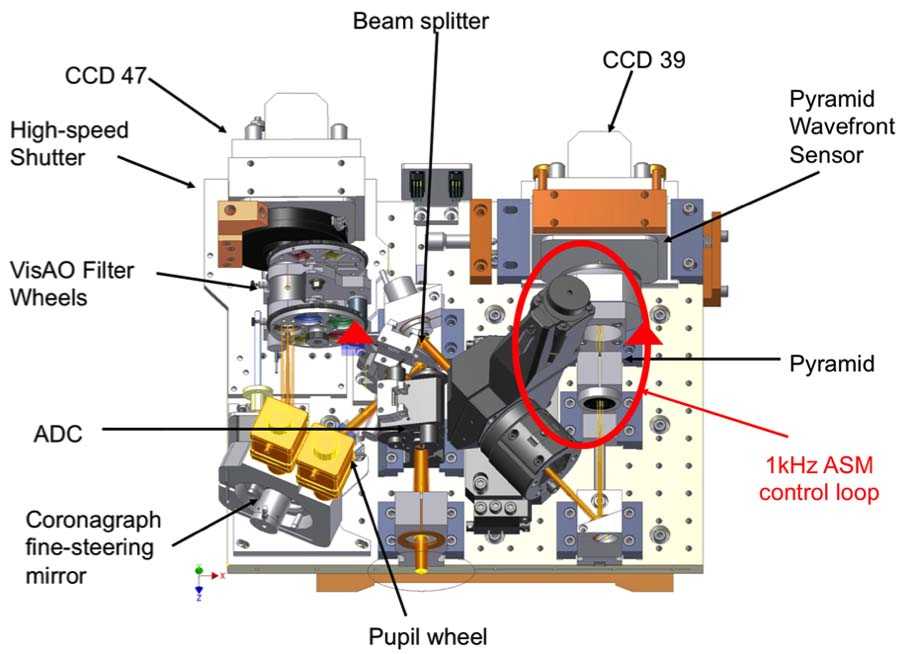}
    \caption{\label{fig:wunit}
    The W-unit containing the PWFS and VisAO.
    Left:
    This view is from the side that is nearer the telescope.
    The infrared light passes straight through the beamsplitter at top left to Clio2 (not shown, would be behind black structure at top left).
    The visible light is split between the PWFS arm (shown in red) and the VisAO arm (shown in blue).
    Right:
    Schematic of the W-unit containing the PWFS and VisAO (somewhat upside-down related to Fig.~\ref{fig:wunit}).
    Here, light enters from M3, is reflected off the Clio2 beamsplitter, and enters at the bottom.
    The triplet ADC corrects atmospheric dispersion before the light is split between PWFS and VisAO via a selectable beamsplitter.
    Then the PWFS light travels to the right, off a PI mirror providing the modulation, through the K-mirror re-rotator that keeps the pupils upright, to the pyramid where the pupil is split into 4 pupil images on the CCD 39.
    The VisAO light travels to the left, off a fine-steering gimbal mirror, through the VisAO filters, to the VisAO CCD 47.
    }
\end{figure}

The pyramid wavefront sensor (PWFS) is co-located with the VisAO science camera (see \S \ref{sec:visao}) on the ``W-unit'' breadboard.  The W-unit is installed on a stiff 3-axis Bayside stage for positioning to micron-level accuracy.  The W-unit is shown in Fig.~\ref{fig:wunit} and described in detail in Close et al.\cite{close2014spie,close2012spie}.
A more detailed schematic view is shown in Fig.~\ref{fig:wunit}, right.  The first beamsplitter is fixed, and transmits the $\lambda>$~1~$\mu$m light to the infrared science camera Clio2, and reflects the shorter wavelength light into the W-unit.  A second beamsplitter is located on the W-unit itself and transmits a selectable portion of the visible light to the PWFS, and reflects the remainder to VisAO.  A triplet atmospheric dispersion compensator (ADC) is common to both the PWFS and VisAO\cite{kopon2013}, while a K-mirror ``re-rotator'' on the PWFS channel keeps the pupils aligned as the telescope tracks.  The W-unit is located inside the Nasmyth (``NAS'') ring, and rotates with the NAS unit.  A PI tip-tilt mirror on a piezo stage is used to modulate the spot on the tip of the pyramid.  The frequency and amplitude of the modulation are parameters chosen for the seeing conditions and guidestar brightness, based on a look-up table calibrated during commissioning.

The diffraction-limited spot at the tip of the pyramid is 30~$\mu$m in diameter under good seeing and bright NGS conditions, while the tip of the glass pyramid itself is 6~$\mu$m across\cite{tozzi2008}.  However, atmospheric dispersion can extend the spot up to 2000~$\mu$m\cite{kopon2008}.  Therefore, a novel triplet ADC was developed\cite{kopon2013} to allow for the MagAO PWFS and VisAO to achieve excellent performance across to a wide range of zenith angles.

The PWFS CCD is an e2v CCD 39 with SciMeasure Little Joe electronics, with 80x80 pixels that are read with four amplifiers.
The frame rates are selectable up to 1053 Hz, and the pixel rate is up to 2500 kHz.
The gains can be varied from $\sim$0.5 e$^-$/ADU up to $\sim$12.5 e$^-$/ADU.
Thus we can adjust the frame rate and gain to lock on a star as bright as 0$^\mathrm{th}$-mag.\ NGS $\alpha$ Cen A (see Males et al.\cite{males2014spie}, these proceedings), and as faint as R$\sim$16$^\mathrm{th}$-mag.\ (see Fig.~\ref{fig:ngs}), without any adjustments to the optics.

The PI mirror is commanded with equal stroke in X and Y, which causes the spot to modulate in a circular pattern centered on the the tip of the pyramid.  The frequency of modulation of the PI mirror is matched to the frame rate of the CCD 39, and the amplitude is selected from a lookup table to adjust the ``optical gain'' for the seeing conditions and number of K-L modes being corrected by the top-level AO loop.  That is, a larger amplitude of modulation is required under poor seeing conditions or for faint guide stars with only a lower-order correction applied, because the spot size is larger and therefore the optical gain (the WFS output related to the phase aberration input) would be reduced if the spot were modulated only a small amount about the tip of the pyramid.

\subsection{Adaptive secondary mirror}
\begin{figure}[!h]
    \centering
    \includegraphics[width=0.7\linewidth]{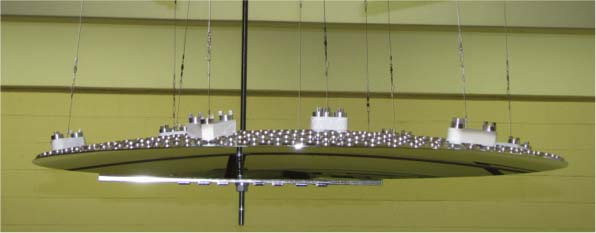}
    \includegraphics[width=0.52\linewidth]{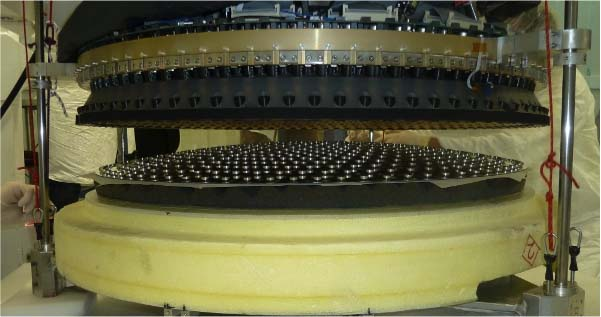}
    \includegraphics[width=0.25\linewidth]{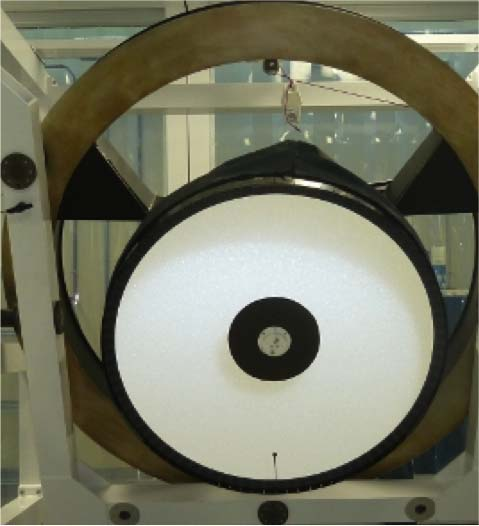}
    \caption{\label{fig:asm}
    The MagAO ASM.
    Top: the thin shell (85.1 cm in diameter, 1.6 mm thick) being lifted out of the aluminizing chamber at Steward Observatory Mirror Lab.
    Bottom left: Inspecting the ASM for damage at the LCO clean room after safe transport to Chile (no damage was found).
    Bottom right: The ASM tilted on its handling cart at the LCO clean room, to conduct operational tests, after safe transport to Chile.
    }
\end{figure}

The wavefront is corrected by a 585-actuator adaptive secondary mirror (ASM), a concave ellipsoid at the Gregorian secondary position of the Clay.  The MagAO ASM is a second-generation mirror, shown in Fig.~\ref{fig:asm}.  The thin shell (TS) is 85.1 cm in diameter and 1.6 mm thick of Zerodur glass.  585 magnets are glued to the back, and they fit into 585 holes cored in the Zerodur reference body, with capacitive sensors that sense the positions of the actuators to 2--5 nm rms position precision.
The position is controlled by a 70 kHz control loop, with a modal settling time $<$~1 ms\cite{close2010spie,riccardi2008spie,riccardi2010spie}.
The ASM thin shell was polished at the Steward Observatory Mirror Lab before the thinning process to an optical quality that, after simulating actuator compensation with negligible force, resulted to be better than 20 nm wavefront error (WFE) rms (Martin et al.\ (2006), Shell A\cite{martin2006spie}).
The shell shape in operation (after thinning, magnet gluing, and actuator compensation) was qualified during the Optical Acceptance Test in Arcetri and resulted in 30 nm rms residual WFE.

\subsection{Software}
The AO subsystems on MagAO use software and firmware essentially identical to those used for the FLAO and LBTI AO systems at the Large Binocular Telescope (LBT), Arizona. As such, the descriptions here are brief and further details can be read in the Arcetri and LBTO documentation for the FLAO and LBTI AO systems, respectively.

The pyramid pupils are imaged on to the CCD 39 which is read out using SciMeasure Little Joe electronics, and processed by the Microgate BCU 39 (basic computing unit) frame grabber.  The BCU 39 calculates slopes, which are then sent via a proprietary fiber optic ``fastlink'' to a cluster of BCUs which perform the reconstruction and control the ASM.  Thus the real-time AO calculations are computed in the BCU control electronics programmed by our partners at Microgate and ADS.  For more details please see Biasi et al.\ (2003)\cite{biasi2003}.

The ASM is commanded via Karhunen-Lo{\`e}ve (K-L) mirror modes which are calculated by our partners at Arcetri and calibrated through measurements of the interaction matrices at the telescope by means of the Calibration Return Optic (CRO, see \S\ref{sec:cro}).   The K-L modes are chosen because they best match the modes of the mirror and the power spectrum of Kolmogorov turbulence. Details of the generation and calibration of the mirror commands are described in Quir{\'o}s-Pacheco et al.\ (2010)\cite{quirospacheco2010}.

The top-level AO software is also an LBT clone, and consists of a supervisor process that monitors the other AO processes as state variables, as described in Fini et al.\ (2008)\cite{fini2008}. The ASM and PWFS are subsystems that are controlled by their own processes and communication protocols between them are rendered with the ``Message Daemon'' diagnostics and telemetry streams between the various hardware, firmware, and software components.

\section{M\lowercase{ag}AO operations}
Because MagAO has its own secondary mirror that is not used for other instruments at Magellan, we currently operate under a model designed to minimize secondary mirror change outs.  Therefore, our science runs are of order 1 month in duration, occuring on the order of twice per year.  The first commissioning run (Comm1) was in Nov.--Dec.\ 2012, the second (Comm2) in Mar.--April 2013, and the first community-wide science run was in Mar.--April 2014.  The second community-wide science run will be in Nov.--Dec.\ 2014.  Installing the secondary mirror and the instrument takes 1--2 days, including engineering night time required to calibrate the active optics.  The MagAO guider is integrated into the NAS and is the only part of MagAO mainained by the Magellan staff at LCO.

\subsection{Calibration}
\label{sec:cro}

\begin{figure}[!h]
	\centering
	\includegraphics[width=0.3\linewidth]{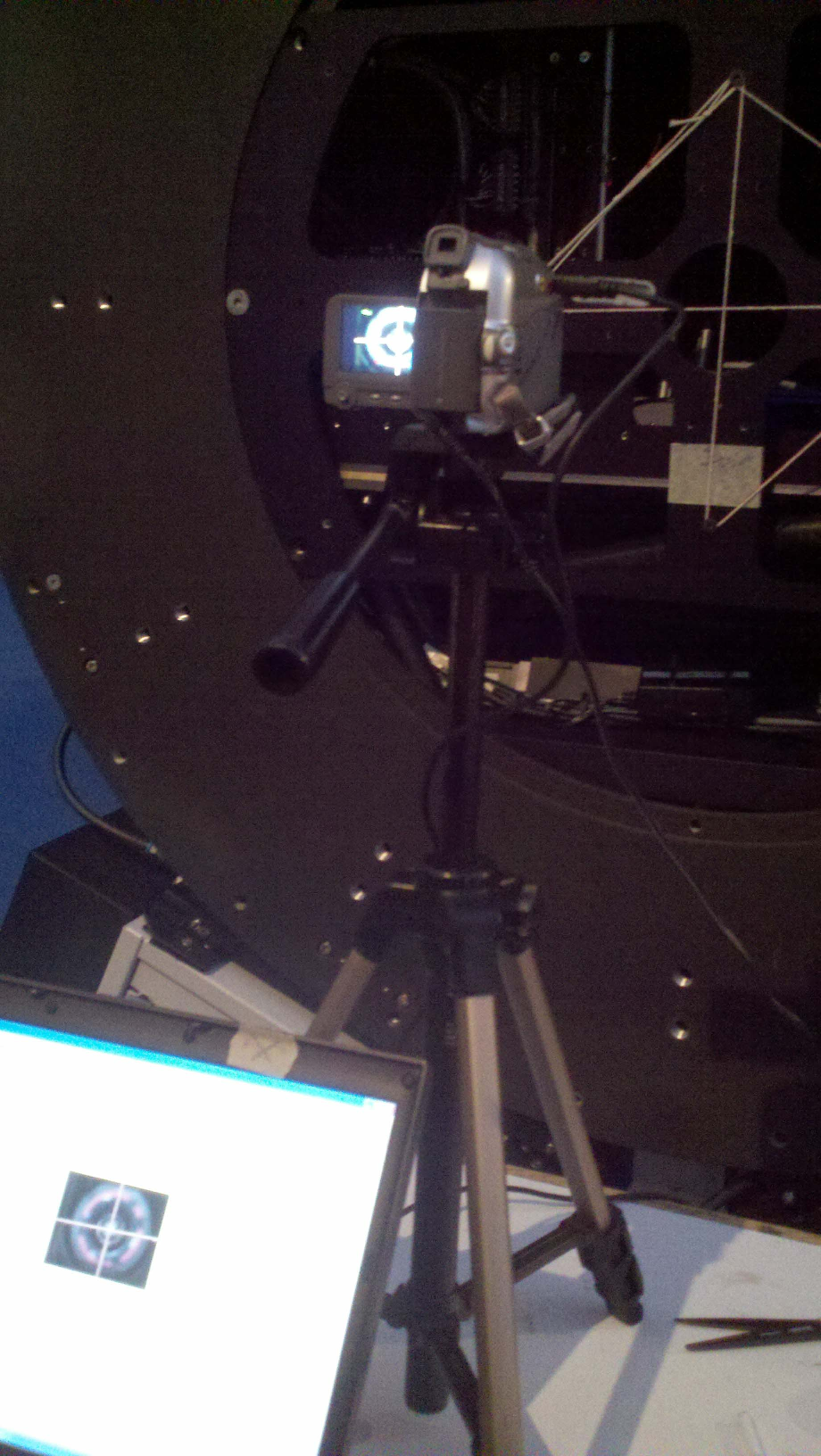}
	\hfill
	\includegraphics[width=0.65\linewidth]{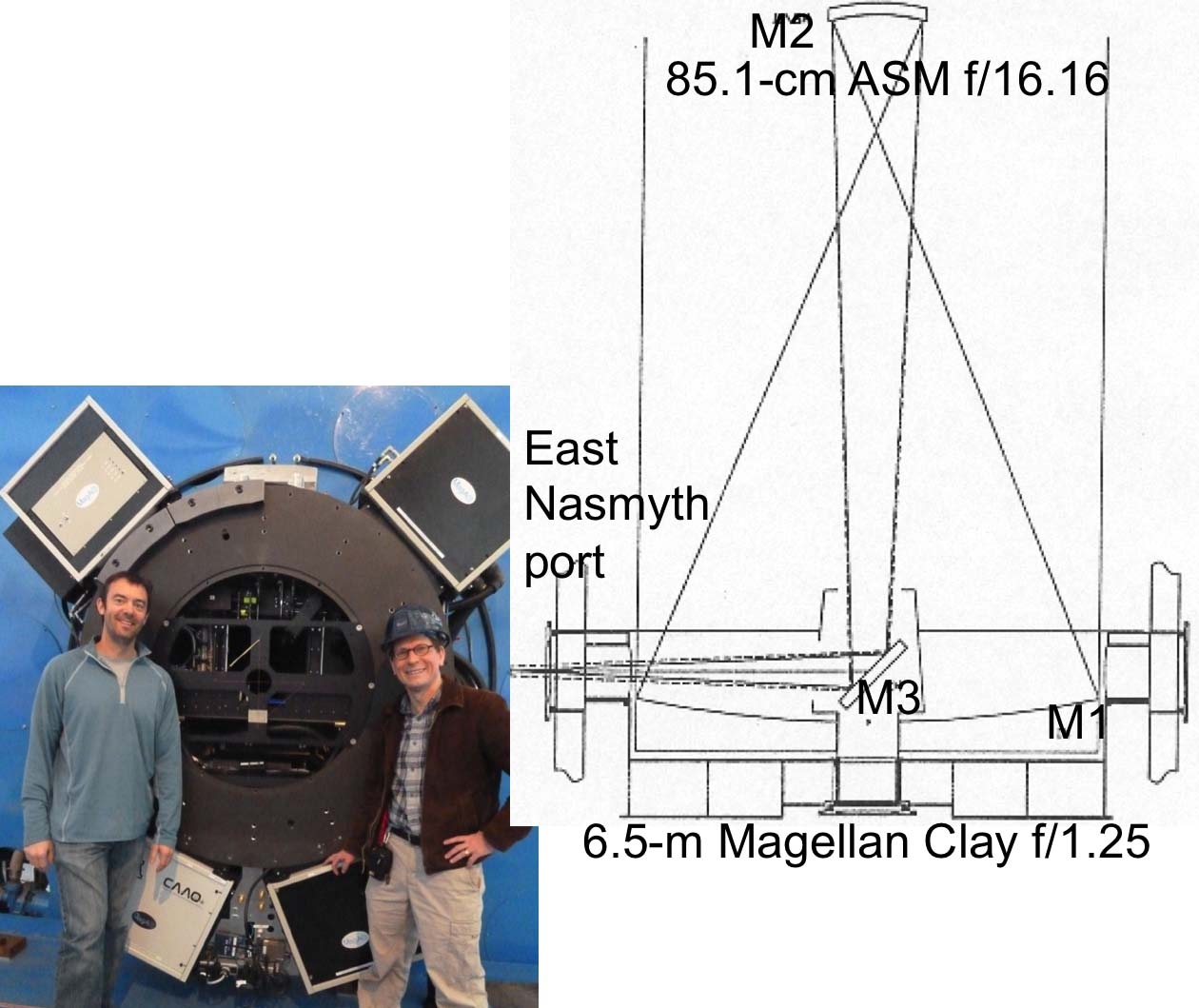}
	\caption{\label{fig:clay_nas_cro}
	Alignment of MagAO.
	At right is a diagram of the optical path, including the intermediate focus located just below the ASM (base diagram from \texttt{http://www.lco.cl/telescopes-information/magellan/}, see Shectman 1994\cite{shectman1994} for more on the Magellan Clay optical design).
	At center is a photo of the W-unit mounted on the NAS, \emph{without} Clio2 so that there is a clear sight to the ASM from the NAS platform.
	At left is a photo of the video camera set-up to align the ASM with the telescope optical axis.
	The NAS ring with a string cross-hair is clearly seen at the center of both photos.
	}
\end{figure}

The MagAO ASM, like the LBT ASMs, is a concave ellipsoid, and thus has an intermediate focus just below it when hanging above the telescope at zenith.  Therefore, we can calibrate the interaction matrices using a calibration source that also samples the whole optical train up to the secondary mirror (excluding the primary mirror).  Figure~\ref{fig:clay_nas_cro} illustrates the set-up.  The calibration return optic (CRO) is a retro-reflector-like optic that sits at the direct focus of the ASM ($\sim$1 m below it) and is illuminated by a laser launched from the W-unit.  The alignment of the ASM to the optical axis of the telescope is done by sighting along the axis from the NAS focus.  Interaction matrices are then taken of each K-L mirror mode with the CRO during daytime or nighttime.  Care is taken not to vibrate the telescope or slam doors in the dome.  The detailed procedure for calibrating the interaction matrices is described in Esposito et al.\ (2010).\cite{esposito2010}

\subsection{Opto-mechanical control loops of MagAO}
\begin{figure}[!h]
	\centering
	\includegraphics[width=0.75\linewidth]{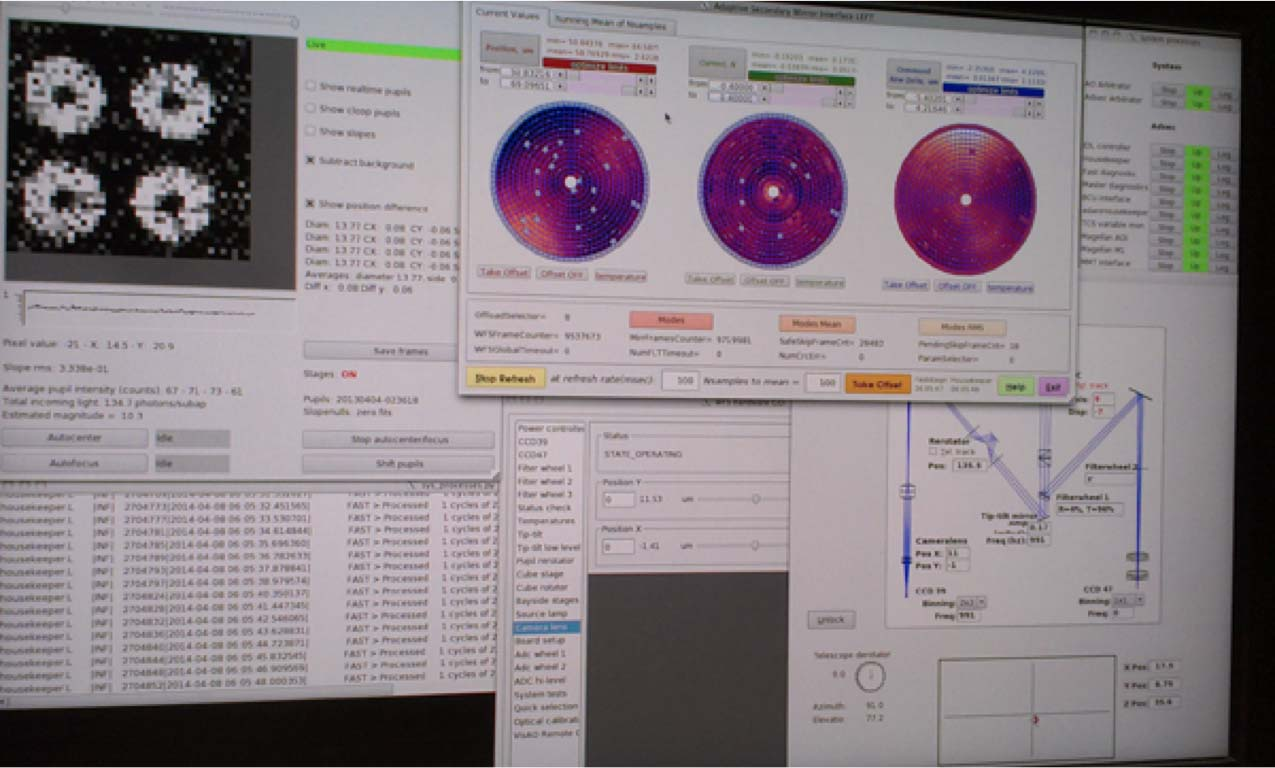}
	\caption{\label{fig:pwfs_asm}
	The PWFS and ASM control GUIs in regular AO operation.
	}
\end{figure}

\begin{figure}[!h]
	\centering
	\includegraphics[width=0.6\linewidth]{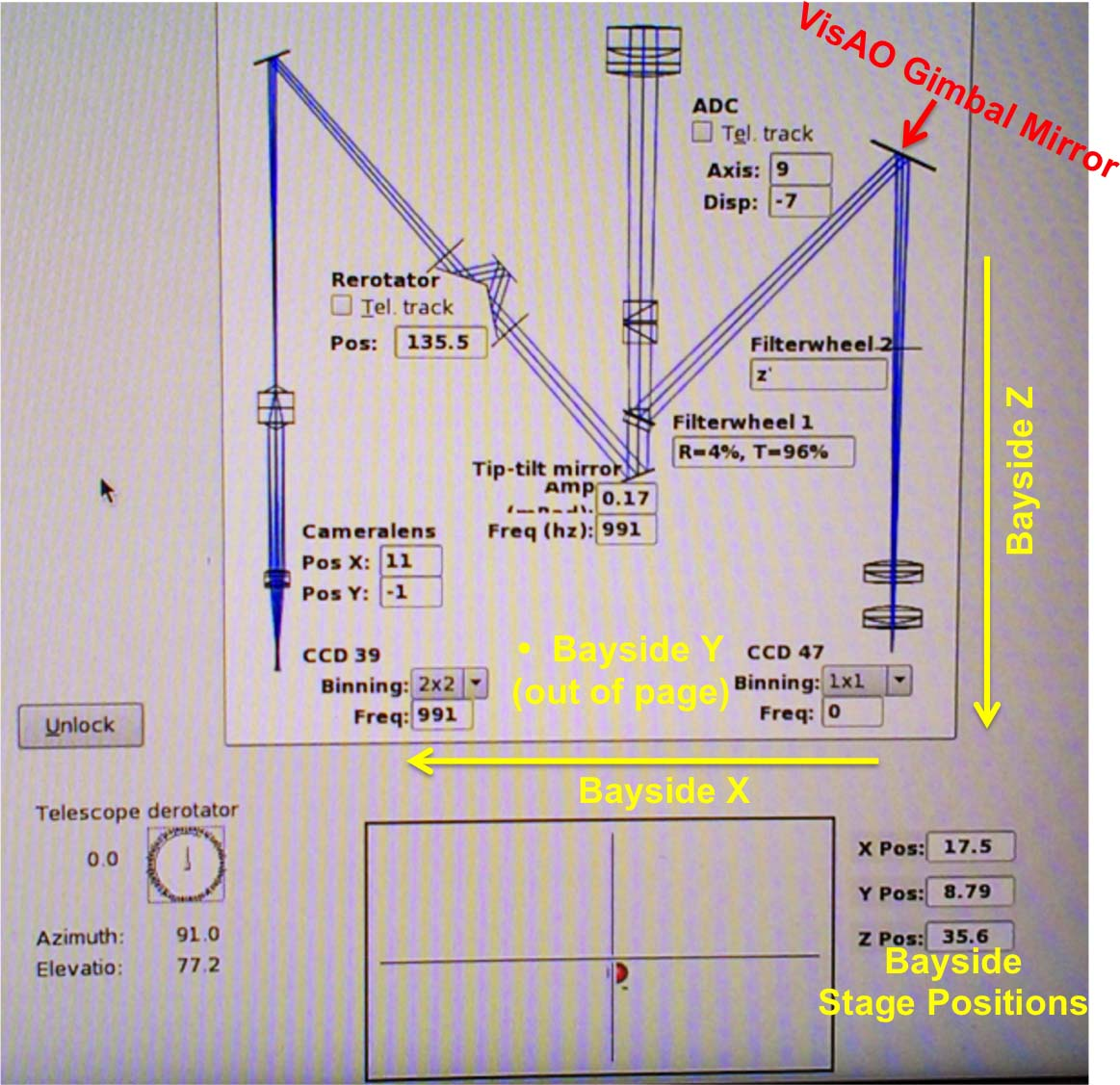}
	\caption{\label{fig:boardgui}
	The W-unit board GUI, showing an overview of the loop status.
	From this diagram, the AO operator can observe:
	the selectable filter chosen to split the visible light between PWFS and VisAO;
	the CCD 39 binning and frequency (which are chosen according to seeing conditions and NGS magnitude);
	the amplitude and modulation frequency of the PI mirror;
	the VisAO science camera parameters such as the science filter and the binning and frame rate of the CCD 47;
	the position of the camera lens as a diagnostic of the camera lens loop;
	the health of the triplet ADC and the re-rotator K-mirror;
	the positions of the bayside stages (which change in X and Y when Clio2 nods and in Z when Clio2 focuses);
	and
	the rotation of the NAS ring.
	}
\end{figure}

As AO systems have matured, they have developed a high level of complexity.
To illustrate, MagAO has several opto-mechanical control loops running at all times.  These are listed here and some are represented in Figs.~\ref{fig:pwfs_asm}--\ref{fig:boardgui}.  Note that this is only a brief highlight, and that many more servo loops are not described here.
\begin{itemize} \itemsep -2pt
\item Top-level AO loop: Runs up to 1 kHz, measuring PWFS slopes, reconstructing phase with the BCU, and sending commands to the ASM.
\item Camera lens loop: Runs at $\sim$1 Hz, keeping the position of the pyramid pupils illuminated on the same pixels on the CCD 39 to a precision of 0.1 pixel.
\item ASM loop: Runs at 70 kHz, with capacative sensors controlling the positions of the actuators to 2--5 nm rms precision.
\item Telescope off-loading: Low-order phase errors measured by the AO system are integrated until they meet a threshold to send to the telescope; a typical frequency is $\sim$1 Hz for tip/tilt, $\sim$0.01 Hz for focus.
\item Active optics: Correction of the primary mirror figure and collimation is done once per new target acquisition, or less often if the slew is small.  A Shack-Hartmann WFS in the MagAO guider is used to measure the correction required.
\item VisAO coronagraph guider (optional): When using the coronagraphic spot on VisAO, this loop measures the position of the ghost and keeps it stable via moving a gimbal mirror, at roughly $\sim$0.1 Hz, in order to keep the star centered behind the coronagraph mask.
\end{itemize}

\subsection{On-sky operations}
The control room has 4 operator stations: the telescope operator (staffed by LCO), the AO operator (staffed by the MagAO team), the Clio2 operator, and the VisAO operator.  The Clio2 and VisAO operators are currently staffed by the MagAO team, but experienced observers can now operate the science cameras themselves, and user manuals and GUIs are being updated to allow this model to become the norm.

The two science cameras, VisAO and Clio2, are separated by a dichroic beamsplitter; therefore, they can be used simultaneously to take visible-light and infrared images of the same star.  Clio2, often operating at thermal wavelengths where sky background is high, must nod as frequently as every few minutes at $M'$-band.  Clio2 nods by sending pointing offsets to the telescope.  These are then counter-acted by the Bayside stages moving to keep the star on the tip of the pyramid, with feedback from the PWFS.  Because the AO loop must briefly pause for the Clio2 nods but the VisAO operator may be unaware of the upcoming pause, VisAO data are time stamped with AO loop status (closed, open, or paused) so that the open-loop frames can be discarded in data reduction.  VisAO integrations are typically 60 seconds or less, so the interuption is usually minimal.  Additionally, we coordinate between the VisAO and Clio2 operators, including prioritizing one camera for each observation, to help determine parameters such as nod frequency.

Visiting astronomers can choose to observe with VisAO, Clio2, or both.  For stars brighter than about $R$$\sim$8, the PWFS receives ample photons from the NGS at 1 kHz such that there is no penalty in AO performance associated with sending half the light to VisAO; thus, VisAO data comes for ``free.''  For stars fainter than this, selecting the bright-star beam splitter may degrade the AO performance by running slower, binning the CCD 39 pixels, and controlling fewer modes; so the astronomer must make a choice about whether to split the visible photons between VisAO and PWFS or to send them all to the AO system.  All IR light ($\lambda$$>$1~$\mu$m) is always being sent to Clio2, so when the astronomer prioritizes VisAO data, they may also obtain Clio2 data if desired.

\section{Science with MagAO}

\subsection{VisAO visible-light AO camera} \label{sec:visao}
The VisAO science camera is an e2v frame transfer CCD 47 with SciMeasure Little Joe electronics, with 1024x1024 pixels that are read with two amplifiers.
The frame grabber is an EDT PCI card.
VisAO is mounted on the W-unit with only a few non-common-path optics from the PWFS, and the two are coupled in motion on the Bayside XYZ stages, so that the resultant image quality on VisAO is closely coupled to the performance of the PWFS.
An exhaustive characterization of the CCD 47 is available in Males (2013).\cite{males2013}

VisAO has two selectable wheels: a filter wheel containing the Sloan filters $r'$, $i'$, $z'$, and $Y_s$ as well as narrow-band filters for spectral differential imaging (SDI) mode.  A Wollaston prism can be inserted to accomplish SDI in H$\alpha$, $[OI]$, and $[SII]$.  Finally, a saturation-suppressing occulting spot of radius 0\farcs1 is available.  A preliminary description of the VisAO photometric system is available in Males (2013)\cite{males2013} and Males et al.\ (2014).\cite{males2014}
See Close et al. (these proceedings\cite{close2014spie}) for a review of the exciting visible wavelength science being done with this system.

\subsection{Clio2 infrared AO camera}
Clio2 (P.I.\ Phil Hinz, UA) has been relocated to Magellan from its initial home at the MMT at Mt.\ Hopkins, Arizona, where it was developed for imaging low-mass companions at 3--5 $\mu$m \cite{clio2004,clio2006,clio2008,clio2010}.  The instrument has two cameras: a wide 28''-field of view camera with 27.5 mas pixels (Nyquist $K$), and a narrow 16''-FoV camera with 15.8 mas pixels (Nyquist $J$).  Exact values are given on our website (\texttt{http://magao.as.arizona.edu}) as they are updated when new calibration data become available.
A suite of IR filters include $J$, $H$, $K_s$, $[3.1\mu\text{m}]$, $[3.3\mu\text{m}]$, $[3.4\mu\text{m}]$, $LÕ$, $[3.9\mu\text{m}]$, $M'$, and two neutral density filters. There is also a low resolution (R$\sim$130 at $3.5\mu\text{m}$; R$\sim$30 at $K$) prism with 0\farcs06, 0\farcs12, and 0\farcs36 slits. Two apodized phase plate (APP) coronagraphs for high-contrast imaging at $L'$ and $M'$, and two non-redundant aperture masks (NRM) complete the observing modes.  However, the APPs are sized for the MMT pupil and the wide camera, and are currently under consideration for an upgrade.

\subsection{Science performance}
MagAO performance has been excellent from the start, due to careful calibration and the excellent base we have built on with the LBT AO systems.  In this section, we present some images illustrating the image quality and AO performance seen on VisAO and Clio2.
See Close et al.\ (these proceedings\cite{close2014spie}) for a review of MagAO visible science.

\begin{figure}[!h]
	\centering
	\includegraphics[width=0.35\linewidth]{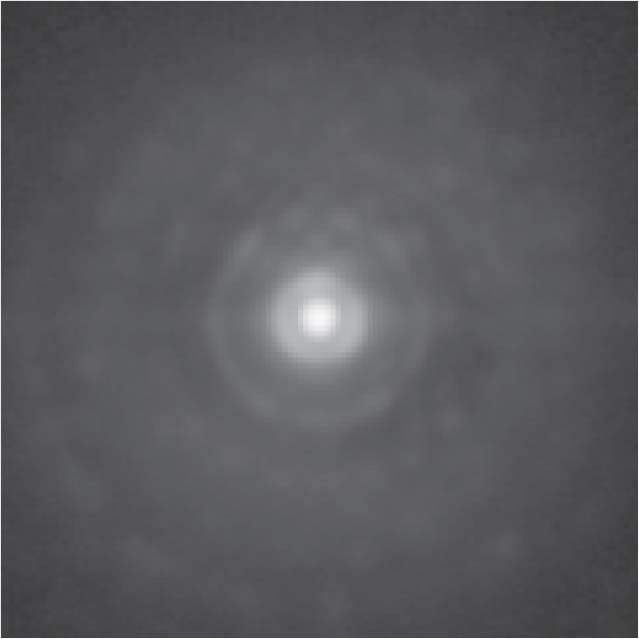}
	\includegraphics[width=0.347\linewidth]{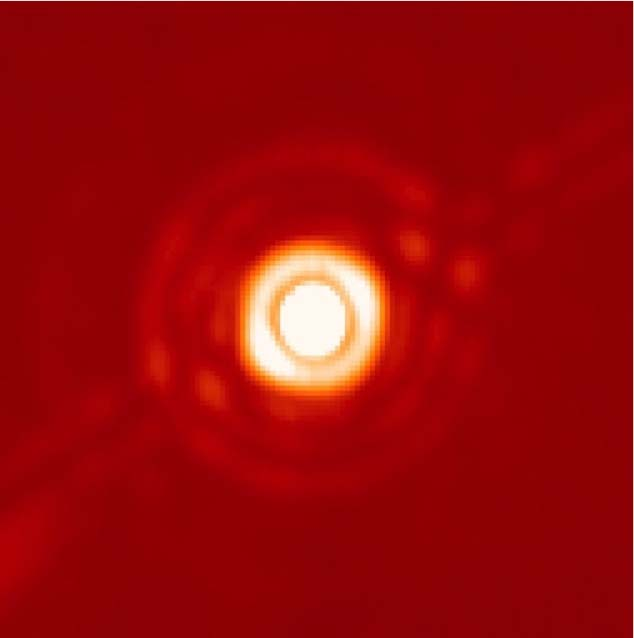}
	\caption{\label{fig:psf}
	PSFs on a bright $R$$\sim$4 guide star with VisAO $Y_s$ (left) and Clio2 $L'$ (right).
	40\% Strehl in $Y_s$ and $>$90\% Strehl at $L'$ for $\sim$137 nm rms WFE.
	}
\end{figure}

Figure~\ref{fig:psf} shows two simultaneous VisAO and Clio2 PSFs on a bright $R$$\sim$4 guide star during Comm1, $\beta$ Pic.  For results on the planet imaged around this star with VisAO and Clio2, see Males et al.\ (2014)\cite{males2014} and Morzinski et al.\ (2014)\cite{morzinski2014}.  We have been able to flatten the wavefront to as good as 125 nm rms WFE on bright guide stars.  After Comm2, we added an upgraded PK50 thermal-blocking filter to improve upon the old BK7 filter to the Clio2 $J$, $H$, and $K_s$ bands, improving our Strehl measurements with Clio2.
See Males et al.\ (these proceedings\cite{males2014spie}) for a review of high contrast work with MagAO.

\begin{figure}[!h]
	\centering
	\includegraphics[width=0.36\linewidth]{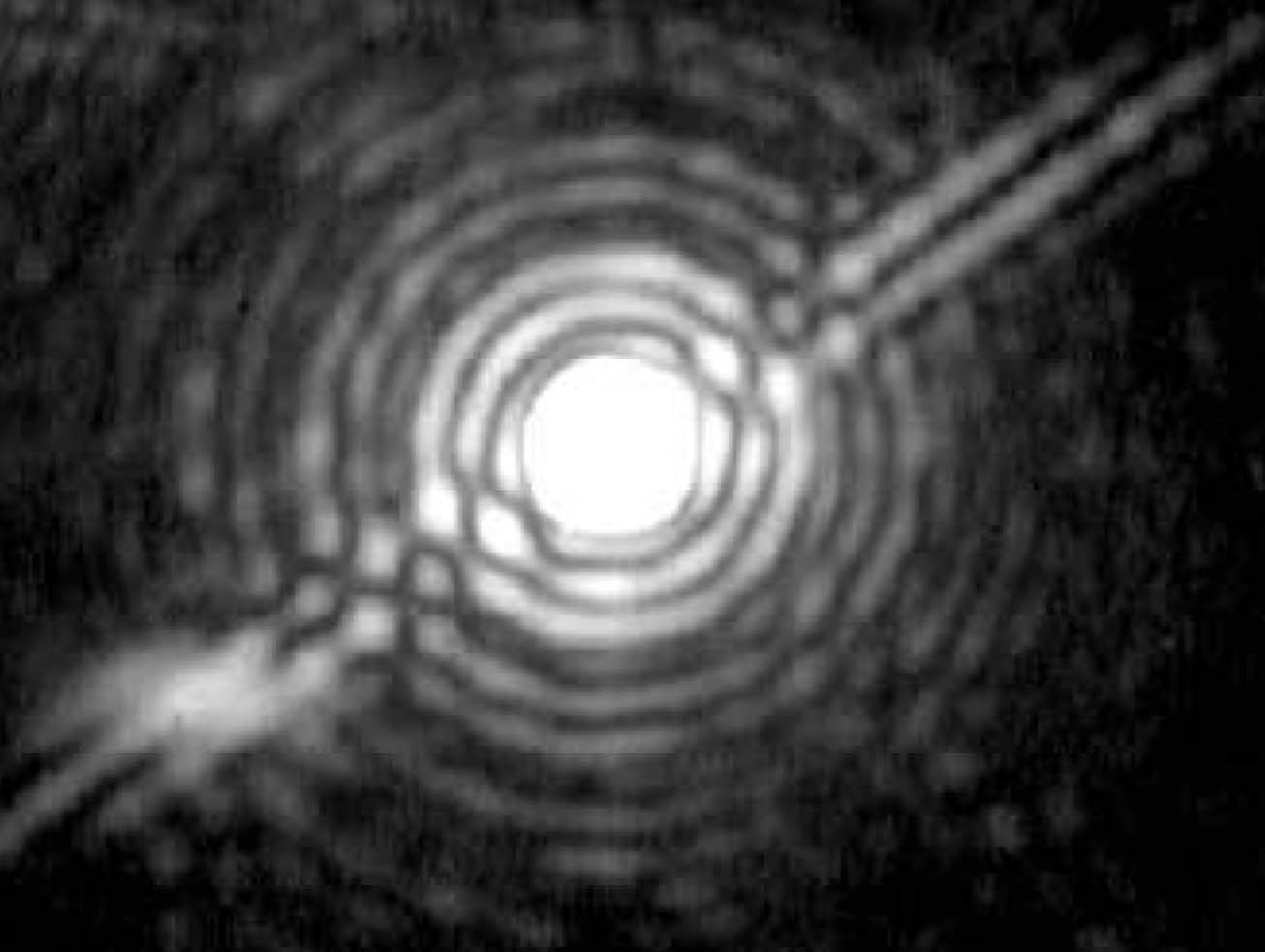}
	\includegraphics[width=0.32\linewidth]{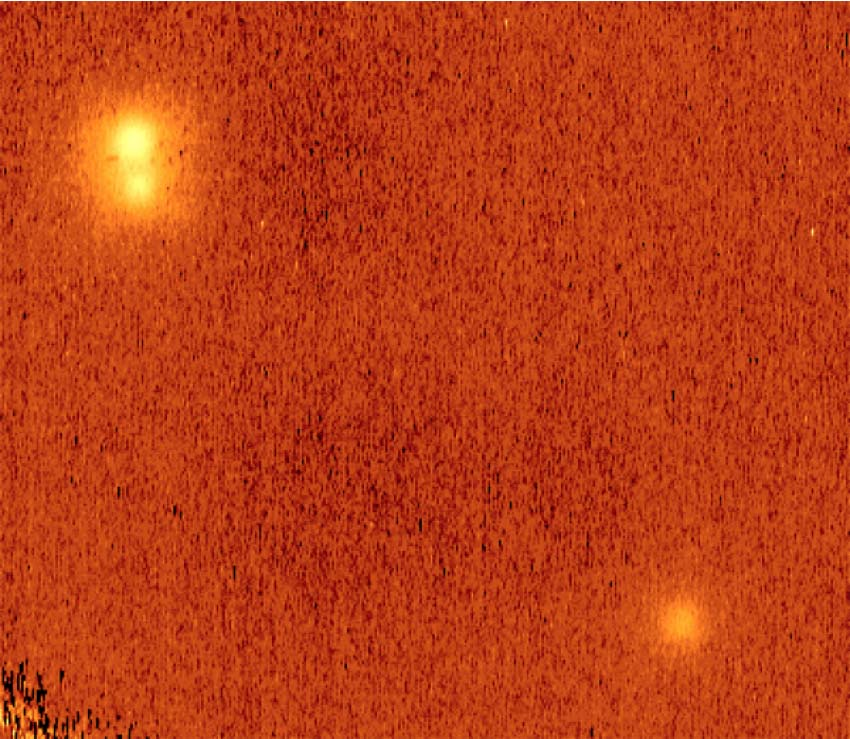}
	\caption{\label{fig:ngs}
	Clio2 images when MagAO is locked on its full range of NGS brightnesses.
	Left: 0$^\mathrm{th}$-mag.\ NGS ($\alpha$ Cen A) at $M'$-band, controlling 378 modes and running at 1 kHz.  See Males et al.\ (these proceedings)\cite{males2014spie} for the results of this observation.
	Right: 15.5$^\mathrm{th}$-mag.\ NGS (upper left binary is guide star) at $K_s$-band, only controlling $\sim$20 modes and running at $\sim$100 Hz.  Star to lower right is about 5'' away from NGS binary and still round.
	}
\end{figure}

Figure~\ref{fig:ngs} shows the performance of MagAO on the brightest and faintest guide stars.

\begin{figure}[!h]
	\centering
	\includegraphics[width=0.85\linewidth]{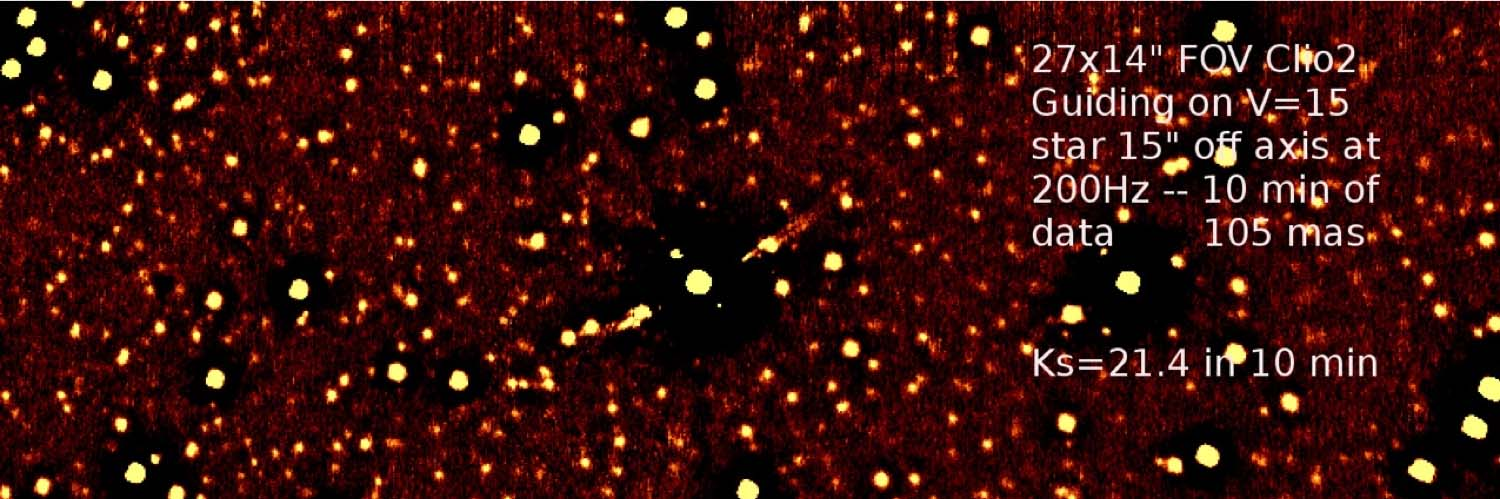}
	\caption{\label{fig:widefield}
	Off-axis wide field correction in $K_s$-band with Clio2.
	The image is unsharp-masked to bring out the many faint stars seen in the field.
	This is the Clio2 wide camera and is about 28'' across.
	Image is cropped in the vertical direction; full wide camera FoV is 28'' x 14''.
	}
\end{figure}

Figure~\ref{fig:widefield} shows our off-axis capabilities.  The guide star is $V$$\sim$15 located 15'' away (to the upper left of the field), running at 200 Hz.  The faintest stars seen in 10 min.\ are $K_s$$\sim$21.4 and the FWHM of the brighter stars are $\sim$105 mas.

\section{Science highlights}
After two month-long commissioning runs and one month-long science run, MagAO users have thus far published 7 ApJ science papers
\cite{close2013,follette2013,wu2013,bailey2014,close2014,males2014,skemer2014}.
Here we highlight a few results.

\begin{figure}[!h]
	\centering
	\includegraphics[width=0.95\linewidth]{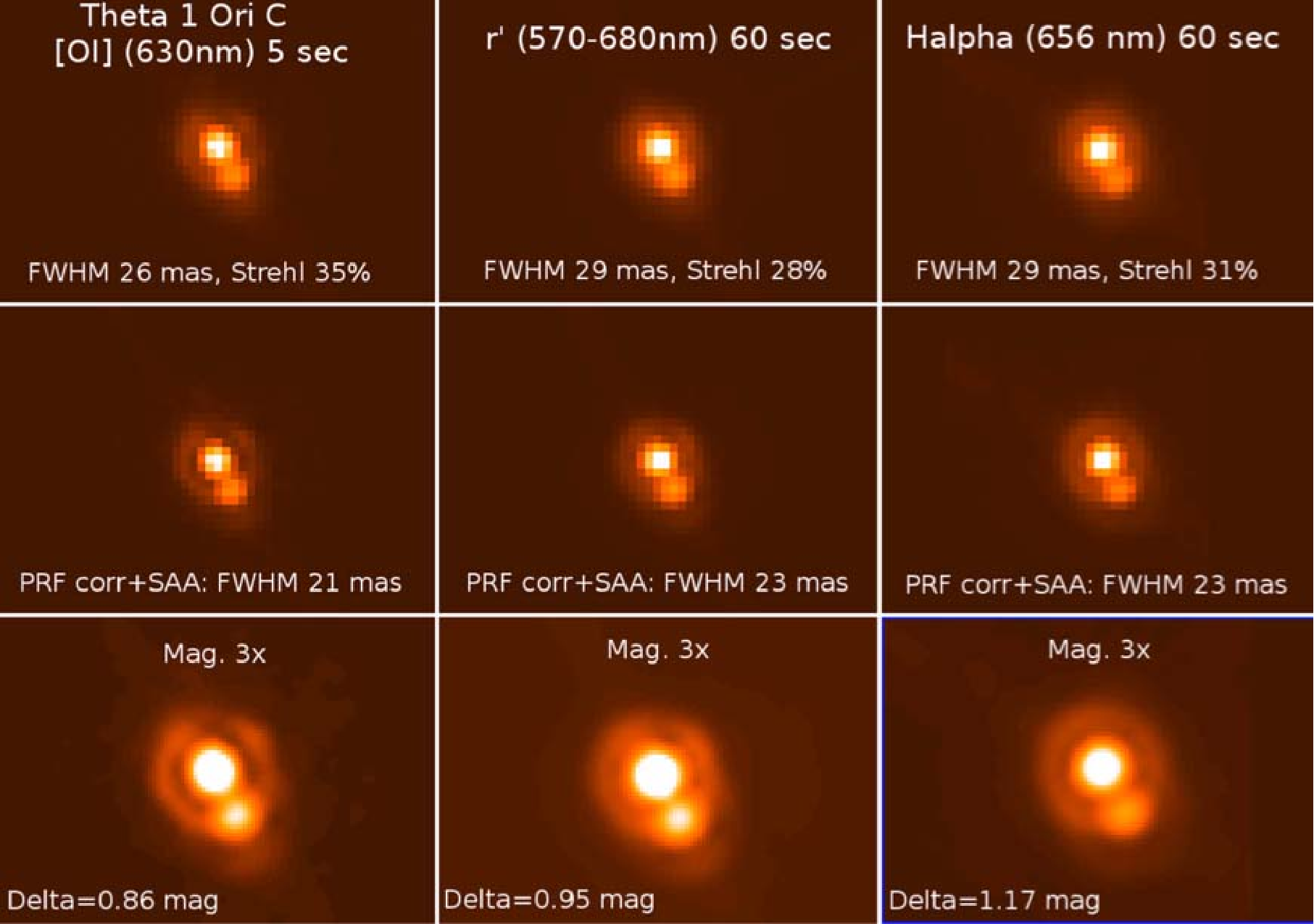}
	\caption{\label{fig:theta1oric}
	VisAO images of $\Theta$ 1 Ori C, a 33-mas binary resolved for the first time with filled-aperture long-exposure direct imaging.
	Figure taken from Close et al.\ (2013).\cite{close2013}
	}
\end{figure}

The massive stars in the Orion Trapezium cluster were used for astrometric calibrations during Comm1.
At the same time, our sharp images were able to resolve, for the first time with filled-aperture long-exposure direct imaging, the binary star $\Theta$ 1 Ori C.
Figure~\ref{fig:theta1oric} shows the exquisite MagAO/VisAO images of this 33 mas binary.
For more VisAO science, please see Close et al.\ (these proceedings\cite{close2014spie}), Follette et al.\ (2013)\cite{follette2013}, Wu et al.\ (2013) \cite{wu2013}, and Males et al.\ (these proceedings\cite{males2014spie}).

\begin{figure}[!h]
	\centering
	\includegraphics[width=0.4\linewidth]{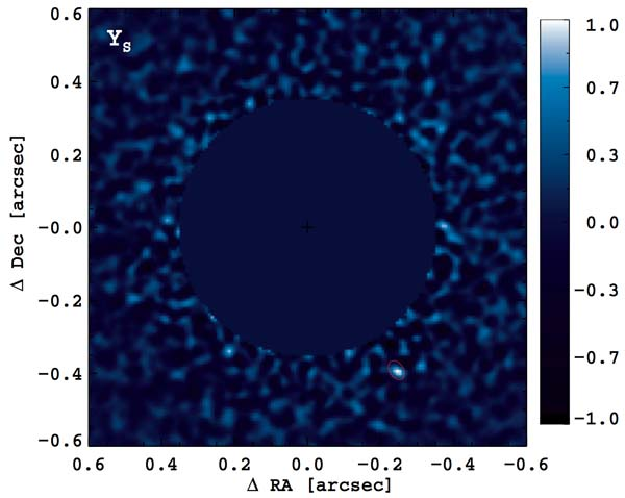}
	\includegraphics[width=0.4\linewidth]{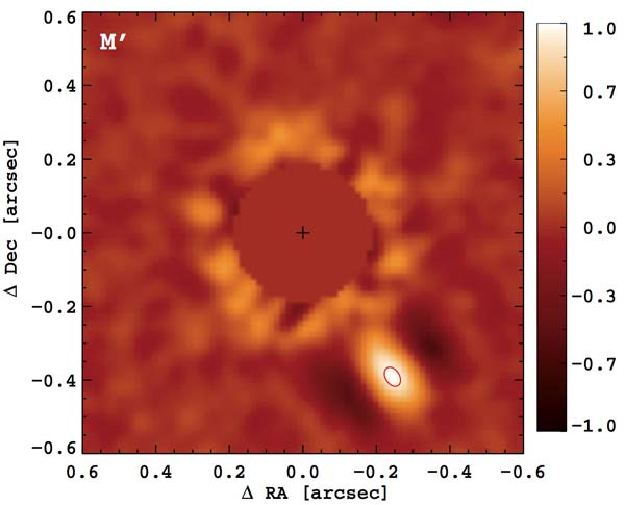}
	\includegraphics[width=0.4\linewidth]{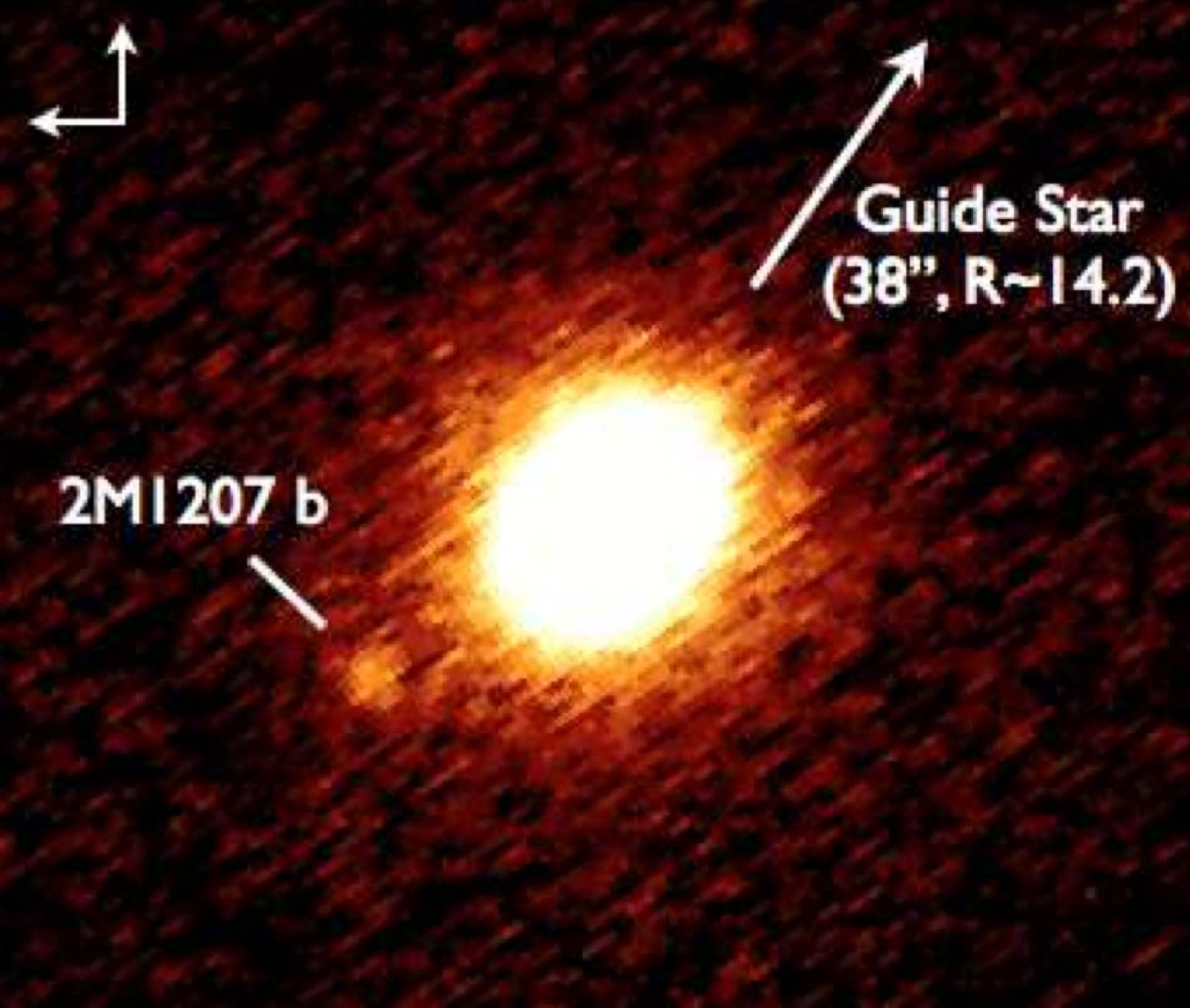}
	\hfill
	\includegraphics[width=0.55\linewidth]{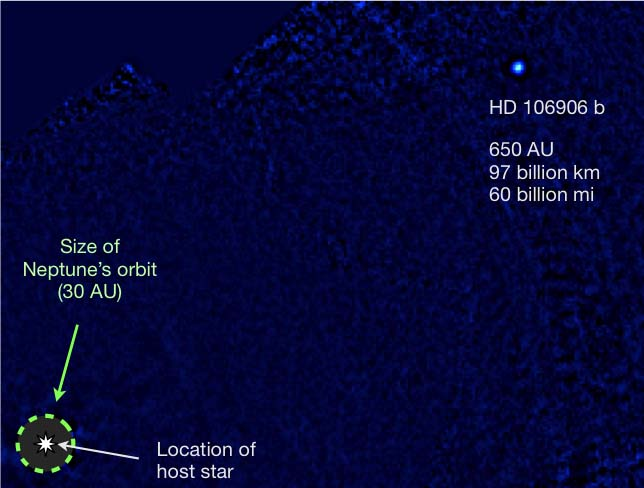}
	\caption{\label{fig:planets}
	Planetary mass companions imaged with MagAO.
	Top:
	MagAO images of extrasolar planet $\beta$ Pic b.
	Top left:
	VisAO $Y_s$ image.
	Top right:
	Clio2 $M'$ image.
	Images taken from Males et al.\ (2014)\cite{males2014}.
	Bottom left:
	Clio2 3.3 $\mu$m image of
	2MASS 1207 b, a planetary-mass companion to a brown dwarf.
	This work also demonstrates the off-axis guiding capabilities.
	Figure taken from Skemer et al.\ (2014)\cite{skemer2014}.
	Bottom right: 
	Clio2 $L'$ image of
	HD 106906b, a MagAO-discovered exoplanet.
	See Bailey et al.\ (2014) for more details.\cite{bailey2014}
	}
\end{figure}

High-order adaptive optics instruments such as MagAO are the best way to directly image planets around nearby stars, due to the small inner working angles required.
Figure~\ref{fig:planets} highlights some of the planets that have been directly imaged with MagAO thus far.
The first AO image of an exoplanet with a CCD was presented in Males et al.\ (2014)\cite{males2014}, using VisAO to image $\beta$ Pictoris b (Fig.~\ref{fig:planets}, top left).
Wavelength coverage across the thermal IR was also obtained with MagAO, by Morzinski et al.\ (2014)\cite{morzinski2014} using Clio2 (Fig.~\ref{fig:planets}, top right).
Skemer et al.\ (2014)\cite{skemer2014} imaged the planetary-mass companion 2MASS 1207 b 
in the thermal infrared methane band at 3.3 $\mu$m with Clio2 (Fig.~\ref{fig:planets}, bottom left).
Finally, Bailey et al.\ (2014)\cite{bailey2014} discovered a new exoplanet, HD 106906 b, with MagAO/Clio2 (Fig.~\ref{fig:planets}, bottom right).

\section{Future plans}
Clio2 upgrades are planned for the near future.  These are installing a new detector with better cosmetics, and acquiring new Apodizing Phase Plate (APP) coronagraphs sized for the MagAO pupil and Clio2 narrow camera.
For VisAO, we will be obtaining new single-substrate SDI filters (like our current second generation H$\alpha$ SDI filters) and selectable aperture stops for SDI and wide-field modes.
We also plan to mitigate minor vibrations from (1) the MagAO CCD cooling pump and (2) offloads to the secondary vane ends.  However, at this time these are quite minor and only affect science when we are attempting 20-mas images as were obtained of Trapezium stars in Comm1.\cite{close2013}
We will also continue to optimize the AO system, with
improved gain estimation and control,
advanced pyramid WFS techniques,
and to adapt the planned LBT upgrades for MagAO ASM and PWFS.

\section{Conclusions}
MagAO is the first ASM-based AO system in the southern hemisphere.
Using our strategy of cloning the successful LBT AO system, we have achieved 125 nm rms WFE on-sky.
MagAO is producing exciting science, with 7 ApJ science papers published from commissioning data alone\cite{close2013,follette2013,wu2013,bailey2014,close2014,males2014,skemer2014}.
We are in full science operation, providing exquisite diffraction-limited images down to 20 mas to the Magellan-wide community.
We have greatly benefited from the AO that came before us, and we continue to have a productive cross-pollination of ideas, tips, and tricks with LBT AO.

\section*{ACKNOWLEDGMENTS}
MagAO was built with support from the NSF MRI, TSIP and ATI programs. LMC's work is supported by NASA Origins program (now XRP) and NSF AAG program.
KMM's and JRM's work is supported by the NASA Exoplanet Science Institute Sagan Fellowship Program.
This work was performed in part under contract with the Jet Propulsion Laboratory and is funded by NASA through the Sagan Fellowship Program under Prime Contract No.\ NAS7-03001.  JPL is managed for the National Aeronautics Space Administration (NASA) by the California Institute of Technology.
Any opinions, findings, and conclusions or recommendations expressed in this publication are those of the authors and do not necessarily reflect the views of the National Aeronautics Space Administration (NASA) or of The California Institute of Technology.

\bibliography{ktmorz_bib_spie2014}

\begin{thebibliography}{10}

\bibitem{shectman2003}
{Shectman}, S.~A. and {Johns}, M., ``{The Magellan Telescopes},'' {\em Proc.
  SPIE}~{\bf 4837},  910--918 (2003).

\bibitem{thomasosip2008}
{Thomas-Osip}, J.~E., {Prieto}, G., {Johns}, M., and {Phillips}, M.~M.,
  ``{{Giant Magellan Telescope} site evaluation and characterization at {Las
  Campanas Observatory}},'' {\em Proc. SPIE}~{\bf 7012},  70121U (2008).

\bibitem{floyd2010}
{Floyd}, D.~J.~E., {Thomas-Osip}, J., and {Prieto}, G., ``{Seeing, Wind, and
  Outer Scale Effects on Image Quality at the {M}agellan Telescopes},'' {\em
  PASP}~{\bf 122},  731--742 (2010).

\bibitem{osip2008}
{Osip}, D.~J., {Phillips}, M.~M., {Palunas}, P., {Perez}, F., and {Leroy}, M.,
  ``{Magellan Telescopes operations 2008},'' {\em Proc. SPIE}~{\bf 7016},
  701609 (2008).

\bibitem{esposito2011spie}
{Esposito}, S., {Riccardi}, A., {Pinna}, E., {Puglisi}, A.,
  {Quir{\'o}s-Pacheco}, F., {Arcidiacono}, C., {Xompero}, M., {Briguglio}, R.,
  {Agapito}, G., {Busoni}, L., {Fini}, L., {Argomedo}, J., {Gherardi}, A.,
  {Brusa}, G., {Miller}, D., {Guerra}, J.~C., {Stefanini}, P., and {Salinari},
  P., ``{Large Binocular Telescope Adaptive Optics System: new achievements and
  perspectives in adaptive optics},'' {\em Proc. SPIE}~{\bf 8149},  814902
  (2011).

\bibitem{close2008spie}
{Close}, L.~M., {Gasho}, V., {Kopon}, D., {Hinz}, P.~M., {Hoffmann}, W.~F.,
  {Uomoto}, A., and {Hare}, T., ``{The Magellan Telescope adaptive secondary AO
  system},'' {\em Proc. SPIE}~{\bf 7015},  70150Y (2008).

\bibitem{close2010spie}
{Close}, L.~M., {Gasho}, V., {Kopon}, D., {Males}, J., {Follette}, K.~B.,
  {Brutlag}, K., {Uomoto}, A., and {Hare}, T., ``{The Magellan Telescope
  Adaptive Secondary AO System: a visible and mid-IR AO facility},'' {\em Proc.
  SPIE}~{\bf 7736},  773605 (2010).

\bibitem{close2012spie}
{Close}, L.~M., {Males}, J.~R., {Kopon}, D.~A., {Gasho}, V., {Follette}, K.~B.,
  {Hinz}, P., {Morzinski}, K., {Uomoto}, A., {Hare}, T., {Riccardi}, A.,
  {Esposito}, S., {Puglisi}, A., {Pinna}, E., {Busoni}, L., {Arcidiacono}, C.,
  {Xompero}, M., {Briguglio}, R., {Quiros-Pacheco}, F., and {Argomedo}, J.,
  ``First closed-loop visible {AO} test results for the advanced adaptive
  secondary {AO} system for the {M}agellan telescope: {MagAO}'s performance and
  status,'' {\em Proc. SPIE}~{\bf 84470},  84470 (2012).

\bibitem{ragazzoni1996}
{Ragazzoni}, R., ``{Pupil plane wavefront sensing with an oscillating prism},''
  {\em Journal of Modern Optics}~{\bf 43},  289--293 (1996).

\bibitem{ragazzoni1999}
{Ragazzoni}, R. and {Farinato}, J., ``{Sensitivity of a pyramidic Wave Front
  sensor in closed loop Adaptive Optics},'' {\em A\&A}~{\bf 350},  L23--L26
  (1999).

\bibitem{close2014spie}
{Close}, L.~M., {Males}, J.~R., Follette, K.~B., Hinz, P., {Morzinski}, K., Wu,
  Y.-L., Kopon, D., Riccardi, A., Esposito, S., Puglisi, A., Pinna, E., ,
  Xompero, M., Briguglio, R., and Quiros-Pacheco, F., ``{Into the Blue: {AO}
  Science with {MagAO} in the Visible},'' {\em Proc. SPIE}~{\bf 9148},  paper
  \#9148--56 (these proceedings) (2014).

\bibitem{kopon2013}
{Kopon}, D., {Close}, L.~M., {Males}, J.~R., and {Gasho}, V., ``{Design,
  Implementation, and On-Sky Performance of an Advanced Apochromatic Triplet
  Atmospheric Dispersion Corrector for the {M}agellan Adaptive Optics System
  and {VisAO} Camera},'' {\em PASP}~{\bf 125},  966--975 (2013).

\bibitem{tozzi2008}
{Tozzi}, A., {Stefanini}, P., {Pinna}, E., and {Esposito}, S., ``{The double
  pyramid wavefront sensor for LBT},'' {\em Proc. SPIE}~{\bf 7015},  701558
  (2008).

\bibitem{kopon2008}
{Kopon}, D., {Close}, L.~M., and {Gasho}, V., ``{An advanced atmospheric
  dispersion corrector for extreme {AO}},'' {\em Proc. SPIE}~{\bf 7015},
  70156M (2008).

\bibitem{males2014spie}
{Males}, J.~R., {Close}, L.~M., {Guyon}, O., {Morzinski}, K., {Puglisi}, A.,
  {Hinz}, P., {Follette}, K.~B., {Monnier}, J., {Tolls}, V., {Rodigas}, T.~J.,
  Weinberger, A., Boss, A., Kopon, D., Wu, Y.-L., Esposito, S., Riccardi, A.,
  Xompero, M., Briguglio, R., and Pinna, E., ``{Direct Imaging of Exoplanets in
  the Habitable Zone with Adaptive Optics},'' {\em Proc. SPIE}~{\bf 9148},
  paper \#9148--69 (these proceedings) (2014).

\bibitem{riccardi2008spie}
{Riccardi}, A., {Xompero}, M., {Zanotti}, D., {Busoni}, L., {Del Vecchio}, C.,
  {Salinari}, P., {Ranfagni}, P., {Brusa Zappellini}, G., {Biasi}, R.,
  {Andrighettoni}, M., {Gallieni}, D., {Anaclerio}, E., {Martin}, H.~M., and
  {Miller}, S.~M., ``{The adaptive secondary mirror for the Large Binocular
  Telescope: results of acceptance laboratory test},'' {\em Proc. SPIE}~{\bf
  7015},  701512 (2008).

\bibitem{riccardi2010spie}
{Riccardi}, A., {Xompero}, M., {Briguglio}, R., {Quir{\'o}s-Pacheco}, F.,
  {Busoni}, L., {Fini}, L., {Puglisi}, A., {Esposito}, S., {Arcidiacono}, C.,
  {Pinna}, E., {Ranfagni}, P., {Salinari}, P., {Brusa}, G., {Demers}, R.,
  {Biasi}, R., and {Gallieni}, D., ``{The adaptive secondary mirror for the
  Large Binocular Telescope: optical acceptance test and preliminary on-sky
  commissioning results},'' {\em Proc. SPIE}~{\bf 7736},  77362C (2010).

\bibitem{martin2006spie}
{Martin}, H.~M., {Brusa Zappellini}, G., {Cuerden}, B., {Miller}, S.~M.,
  {Riccardi}, A., and {Smith}, B.~K., ``{Deformable secondary mirrors for the
  LBT adaptive optics system},'' {\em Proc. SPIE}~{\bf 6272},  62720U (2006).

\bibitem{biasi2003}
{Biasi}, R., {Andrighettoni}, M., {Veronese}, D., {Biliotti}, V., {Fini}, L.,
  {Riccardi}, A., {Mantegazza}, P., and {Gallieni}, D., ``{{LBT} adaptive
  secondary electronics},'' {\em Proc. SPIE}~{\bf 4839},  772--782 (2003).

\bibitem{quirospacheco2010}
{Quir{\'o}s-Pacheco}, F., {Busoni}, L., {Agapito}, G., {Esposito}, S., {Pinna},
  E., {Puglisi}, A., and {Riccardi}, A., ``{First light {AO} ({FLAO}) system
  for {LBT}: performance analysis and optimization},'' {\em Proc. SPIE}~{\bf
  7736},  77363H (2010).

\bibitem{fini2008}
{Fini}, L., {Tosetti}, F., {Busoni}, L., {Puglisi}, A., and {Xompero}, M.,
  ``{The {LBT}-{AdOpt} arbitrator: coordinating many loosely coupled
  processes},'' {\em Proc. SPIE}~{\bf 7019},  70190F (2008).

\bibitem{esposito2010}
{Esposito}, S., {Riccardi}, A., {Quir{\'o}s-Pacheco}, F., {Pinna}, E.,
  {Puglisi}, A., {Xompero}, M., {Briguglio}, R., {Busoni}, L., {Fini}, L.,
  {Stefanini}, P., {Brusa}, G., {Tozzi}, A., {Ranfagni}, P., {Pieralli}, F.,
  {Guerra}, J.~C., {Arcidiacono}, C., and {Salinari}, P., ``{Laboratory
  characterization and performance of the high-order adaptive optics system for
  the {Large Binocular Telescope}},'' {\em Applied Optics}~{\bf 49},  G174
  (2010).

\bibitem{shectman1994}
{Shectman}, S.~A., ``{Optical design of the Magellan Project 6.5-meter
  telescope},'' {\em Proc. SPIE}~{\bf 2199},  558--564 (1994).

\bibitem{males2013}
Males, J.~R., ``{Towards the habitable zone: Direct imaging of extrasolar
  planets with the Magellan AO system},'' {\em Dissertation}~{\bf University of
  Arizona} (2013).

\bibitem{males2014}
{Males}, J.~R., {Close}, L.~M., {Morzinski}, K.~M., {Wahhaj}, Z., {Liu}, M.~C.,
  {Skemer}, A.~J., {Kopon}, D., {Follette}, K.~B., {Puglisi}, A., {Esposito},
  S., {Riccardi}, A., {Pinna}, E., {Xompero}, M., {Briguglio}, R., {Biller},
  B.~A., {Nielsen}, E.~L., {Hinz}, P.~M., {Rodigas}, T.~J., {Hayward}, T.~L.,
  {Chun}, M., {Ftaclas}, C., {Toomey}, D.~W., and {Wu}, Y.-L., ``{Magellan
  Adaptive Optics First-light Observations of the Exoplanet {$\beta$} Pic {b}.
  I. Direct Imaging in the Far-red Optical with {MagAO}+{VisAO} and in the
  Near-IR with {NICI}},'' {\em ApJ}~{\bf 786},  32 (2014).

\bibitem{clio2004}
{Freed}, M., {Hinz}, P.~M., {Meyer}, M.~R., {Milton}, N.~M., and {Lloyd-Hart},
  M., ``{Clio: a 5-{$\mu$}m camera for the detection of giant exoplanets},''
  {\em Proc. SPIE}~{\bf 5492},  1561--1571 (2004).

\bibitem{clio2006}
{Sivanandam}, S., {Hinz}, P.~M., {Heinze}, A.~N., {Freed}, M., and
  {Breuninger}, A.~H., ``{Clio: a 3-5 micron AO planet-finding camera},'' {\em
  Proc. SPIE}~{\bf 6269},  62690U (2006).

\bibitem{clio2008}
{Hastie}, M. and {McLeod}, B., ``{Comprehensive review of the converted MMT's
  instrument suite},'' {\em Proc. SPIE}~{\bf 7014},  70140B (2008).

\bibitem{clio2010}
{Hastie}, M. and {Williams}, G.~G., ``{Instrumentation suite at the MMT
  Observatory},'' {\em Proc. SPIE}~{\bf 7735},  773507 (2010).

\bibitem{morzinski2014}
Morzinski, K.~M., Males, J.~R., Skemer, A.~J., Close, L.~M., Hinz, P.~M.,
  Puglisi, A., Esposito, S., Riccardi, A., Pinna, E., Xompero, M., Briguglio,
  R., Follette, K., Kopon, D., Gasho, V., Uomoto, A., Hare, T., Arcidiacono,
  C., Quiros-Pacheco, F., Argomedo, J., Busoni, L., Rodigas, T.~J., and Wu,
  Y.-L., ``{Magellan AO First-Light Observations of Beta Pictoris b. II. 3-5
  $\mu$m direct imaging with {MagAO}/Clio2 and SED of a young giant planet from
  0.9-5 $\mu$m},'' {\em {ApJ in prep}}  (2014).

\bibitem{close2013}
{Close}, L.~M., {Males}, J.~R., {Morzinski}, K., {Kopon}, D., {Follette}, K.,
  {Rodigas}, T.~J., {Hinz}, P., {Wu}, Y.-L., {Puglisi}, A., {Esposito}, S.,
  {Riccardi}, A., {Pinna}, E., {Xompero}, M., {Briguglio}, R., {Uomoto}, A.,
  and {Hare}, T., ``{Diffraction-limited Visible Light Images of Orion
  Trapezium Cluster with the Magellan Adaptive Secondary Adaptive Optics System
  {(MagAO)}},'' {\em ApJ}~{\bf 774},  94 (2013).

\bibitem{follette2013}
{Follette}, K.~B., {Close}, L.~M., {Males}, J.~R., {Kopon}, D., {Wu}, Y.-L.,
  {Morzinski}, K.~M., {Hinz.}, P., {Rodigas}, T.~J., {Puglisi}, A., {Esposito},
  S., {Riccardi}, A., {Pinna}, E., {Xompero}, M., and {Briguglio}, R., ``{The
  First Circumstellar Disk Imaged in Silhouette at Visible Wavelengths with
  Adaptive Optics: {MagAO} Imaging of Orion 218-354},'' {\em ApJL}~{\bf 775},
  L13 (2013).

\bibitem{wu2013}
{Wu}, Y.-L., {Close}, L.~M., {Males}, J.~R., {Follette}, K., {Morzinski}, K.,
  {Kopon}, D., {Rodigas}, T.~J., {Hinz}, P., {Puglisi}, A., {Esposito}, S.,
  {Pinna}, E., {Riccardi}, A., {Xompero}, M., and {Briguglio}, R., ``{High
  Resolution H{$\alpha$} Images of the Binary Low-mass Proplyd LV 1 with the
  Magellan AO System},'' {\em ApJ}~{\bf 774},  45 (2013).

\bibitem{bailey2014}
{Bailey}, V., {Meshkat}, T., {Reiter}, M., {Morzinski}, K., {Males}, J., {Su},
  K.~Y.~L., {Hinz}, P.~M., {Kenworthy}, M., {Stark}, D., {Mamajek}, E.,
  {Briguglio}, R., {Close}, L.~M., {Follette}, K.~B., {Puglisi}, A., {Rodigas},
  T., {Weinberger}, A.~J., and {Xompero}, M., ``{HD 106906 b: A Planetary-mass
  Companion Outside a Massive Debris Disk},'' {\em ApJL}~{\bf 780},  L4 (2014).

\bibitem{close2014}
{Close}, L.~M., {Follette}, K.~B., {Males}, J.~R., {Puglisi}, A., {Xompero},
  M., {Apai}, D., {Najita}, J., {Weinberger}, A.~J., {Morzinski}, K.,
  {Rodigas}, T.~J., {Hinz}, P., {Bailey}, V., and {Briguglio}, R., ``{Discovery
  of H{$\alpha$} Emission from the Close Companion inside the Gap of
  Transitional Disk HD 142527},'' {\em ApJL}~{\bf 781},  L30 (2014).

\bibitem{skemer2014}
Skemer, A.~J., Marley, M.~S., Hinz, P.~M., Morzinski, K.~M., Skrutskie, M.~F.,
  Leisenring, J.~M., Close, L.~M., Saumon, D., Bailey, V.~P., Briguglio, R.,
  Defrere, D., Esposito, S., Follette, K.~B., Hill, J.~M., Males, J.~R.,
  Puglisi, A., Rodigas, T.~J., and Xompero, M., ``{Directly Imaged L-T
  Transition Exoplanets in the Mid-Infrared},'' {\em ApJ in press} ,
  arXiv:1311.2085 (2014).

\end{thebibliography}
\bibliographystyle{spiebib}

\end{document}